\documentclass[aps,prb,amsmath,amssymb,twocolumn]{revtex4}
\usepackage{bm}
\usepackage{epsfig}
\usepackage{color}  
\newcommand{\nix}[1]{}
\begin{document}
\title{Optical and photogalvanic properties of graphene supperlattices formed by periodic strain}
\author{Yu.\,Yu.\,Kiselev}
\author{L.\,E.\,Golub}
\email{golub@coherent.ioffe.ru}
\affiliation{Ioffe Physical-Technical Institute of the Russian Academy of Sciences, 194021 St.~Petersburg, Russia}%

\begin{abstract}

Graphene superlattices formed by periodic strain are considered theoretically.
It is shown that electron energy spectrum consists of minibands obtained by folding of the cone at the boundaries of the superlattice Brillouin zone with very narrow anticrossing regions.
Light absorption under direct interminiband transitions 
is shown to be frequency dependent and dichroic. Giant dichroism of absorption is demonstrated for doped  graphene superlattices. 
Asymmetrical graphene superlattices act as quantum ratchets allowing for generation of photocurrent at absorption of normally incident light. The helicity-dependent photocurrent spectrum is calculated for doped superlattices with various asymmetries.
\end{abstract}
\pacs{}

\maketitle

\section{Introduction}

Graphene attracts a great deal of attention due to both interesting physical phenomena and device applications.
Physics of graphene possesses many intriguing effects in optics, transport and other fields.~\cite{Neto09,Peres10} 
An example is optical absorbance of graphene, which is a fundamental constant in a very wide frequency range.
After focusing on pure graphene, the interest is shifted to graphene-based systems whose electronic properties can be changed by external parameters. One of the promising avenues is strain engineering of electronic bandstructure.~\cite{Hamiltonian_strain,Guinea10b} For example, gaps in energy spectrum open in strained graphene.~\cite{Guinea08,Low11} 
In graphene, the strain  is equivalent to gauge, or pseudomagnetic fields,~\cite{Vozmediano10,Guinea10a} which
suppress anomalous magnetoresistance caused by weak localization.~\cite{WL_exp}

Deformation of graphene results in periodic ripples with height $\sim 0.1 \ldots 1$~nm and a period $\sim 10$~nm.~\cite{Parga10,Wang11} Such periodically strained  graphene samples have $D_{2h}$ point symmetry which allows for dichroism of light absorption. Asymmetrical strain lowers symmetry of graphene to $C_{2v}$ with the $C_2$ axis along the strain direction. In such a system a ratchet effect is possible which means a generation of electric current under light absorption of unbiased sample. In particular, a contribution to the photocurrent sensitive to the  helicity of light normally incident on a strained graphene structure is allowed. Up to now such circular ratchet effect has been detected 
only in small graphene samples owing to scattering from sample edges,~\cite{Karch10_condmat} 
or for oblique incidence due to the presence of a substrate.~\cite{Jiang11}

In this work we investigate electronic states and calculate dichroic absorption in periodically strained graphene samples. For asymmetrical periodic strain, we study the circular ratchet effect and calculate the circular photocurrent for various asymmetries of the strain.

\section{Electronic states in graphene superlattices}

\begin{figure*}[t]
\includegraphics[width=0.8\linewidth]{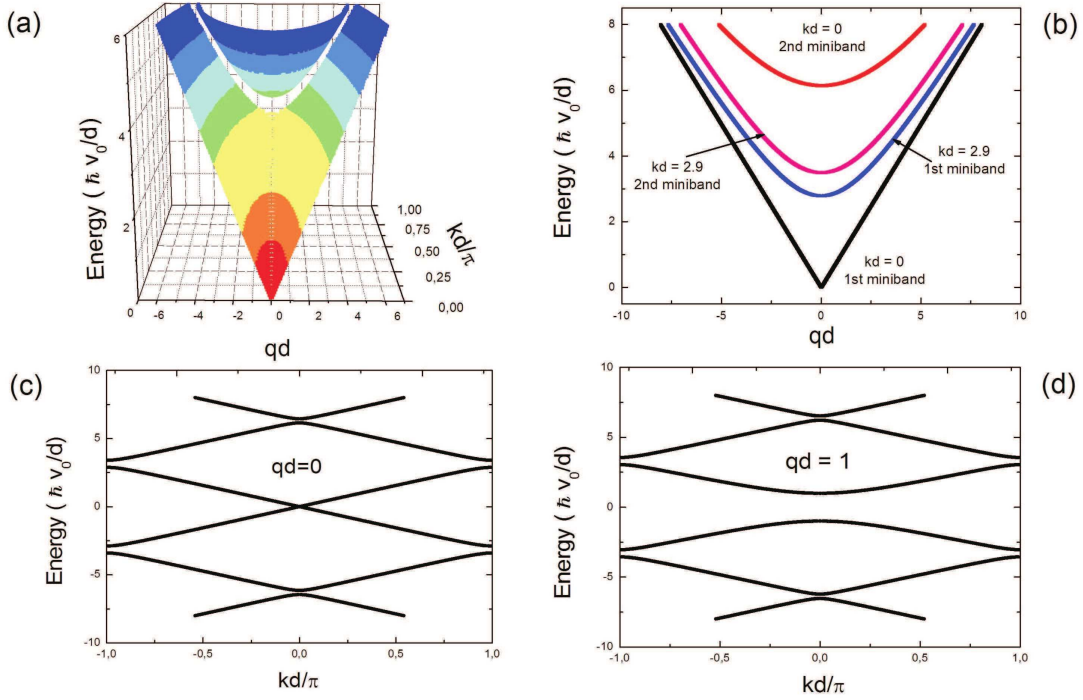}
\caption{Energy spectrum in a SL with a period formed by two layers with widths $l_1=0.3d$, $l_2=0.7d$, and the strain amplitudes  $\gamma_1=0.7$, $\gamma_2=-0.3$.
Three-dimensional plot of the spectrum (a), 
energy dispersion as a function of $q$ for four selected values of $k$ (b), dispersion along $k$ for $q=0$ (c) and $q=1/d$ (d).}
\label{fig1}
\end{figure*}

The Hamiltonian for strained graphene is given by~\cite{Hamiltonian_strain}
\begin{equation}
	H(\bm p) =  v_0 \bm \sigma \cdot (\bm p \pm \bm {\mathcal A}),
\end{equation}
where $\bm p$ is the quasimomentum operator, $\bm \sigma = (\sigma_x,\sigma_y)$ is a vector of Pauli matrices, and the two signs correspond to two valleys.
We consider a graphene sheet periodically strained in the $y$ direction: 
\[
{\mathcal A}_x(y+d)={\mathcal A}_x(y),
\] 
where $d$ is the period. The value of ${\mathcal A}_x$ is modulated with the amplitude $\gamma\hbar/d$ with the estimates~\cite{Parga10} yielding $\gamma \sim 0.03 \ldots 0.3$. Owing to the periodicity, a superlattice (SL) is formed, and  the envelope function satisfies the Bloch theorem
$$
\Psi(x,y+d)=\Psi(x,y){\rm e}^{{\rm i}k d},
$$
where $k$ is the wavevector for motion along the $y$ direction. 

We consider the Kronig-Penney model where strain is a piecewise-constant periodic function of $y$. The wavefunctions within the regions with the fixed value of the strain ${\mathcal A}_x = \gamma_i\hbar/d$ can be cast as
\begin{equation}
\label{psi}
	\Psi_i(x,y)= {\rm e}^{{\rm i}q x} 
	\left[ 
	C_i 
	\left( 
	\begin{array}{c}
	1\\
	{\rm e}^{{\rm i}\varphi_i}
	\end{array}
	\right)
	{\rm e}^{{\rm i}k_i y}
	+
	D_i
	\left( 
	\begin{array}{c}
	1\\
	{\rm e}^{-{\rm i}\varphi_i}
	\end{array}
	\right)
	{\rm e}^{-{\rm i}k_i y}
	\right].
\end{equation}
Here $q$ is the wavevector for free motion along the $x$ direction, $k_i$ are determined by the energy $E$ via $k_i=\sqrt{(E/\hbar v_0)^2 - (q-\gamma_i/d)^2}$, and $\tan{\varphi_i} = \pm k_i/(q-\gamma_i/d)$, where the upper and lower signs should be taken for conduction and valence band states, respectively.
The dispersion equation has the form
\begin{equation}
\label{disp_eq}
	\cos{kd} = (T_{11}+T_{22})/2,
\end{equation}
where $T$ is the transfer matrix. For the periodic structure with $N$ layers in each period it equals to a product $T=T_{N-1}T_{N-2}\ldots T_{1}T_0$, where $T_i$ is the transfer matrix for $i$th strained graphene layer of the width $l_i$. It is defined as $\Psi_i(x,y+l_i)= T_i \Psi_i(x,y)$ and given by
\begin{equation}
	T_i = {1\over \sin{\varphi_i}}
	\left[
		\begin{array}{cc}
	\sin{(\varphi_i - k_i l_i)} & \sin{k_i l_i}\\
	-\sin{k_i l_i}  & \sin{(\varphi_i + k_i l_i)}
	\end{array}
\right].
\end{equation}
For a graphene SL with a period formed by two layers, the dispersion equation has the form~\cite{Dirac-Kronig-Penney}
\begin{eqnarray}
	\cos{kd}  &=& \cos{k_1 l_1} \cos{k_2 l_2} \\
&&	+ \sin{k_1 l_1} \sin{k_2 l_2} {\cos{\varphi_1} \cos{\varphi_2} -1 \over \sin{\varphi_1} \sin{\varphi_2}}
	\nonumber.
\end{eqnarray}
Energy spectrum represents a result of folding of Dirac cone within one-dimensional Brillouin zone of the SL. 
%
%
The cone is shifted from $q=0$ by a value $q_0=(\gamma_1 l_1+\gamma_2 l_2)/d^2$. 
The spectrum is given by conduction minibands $E_{n}(k,q)$ and valence minibands with the inverted dispersion $-E_{n}(k,q)$ with $n=0,1,2\ldots$
The conduction- and valence-band cones touch each other at $q=q_0$, $k=0$, i.e. the zero-energy state is degenerate. 
At the points $k=0,\pm\pi/d$,
the gaps between the minibands are opened.
At small deformation $|\gamma_1-\gamma_2| \ll 1$, the gaps
are given by
\begin{equation}
	E_{n+1}-E_{n}= \frac{2|\gamma_1-\gamma_2| \hbar v_0}{n\pi d} \left| \sin{\frac{n\pi l_1}{d}} \right|.
\end{equation}
 
At arbitrary values of the strain in the layers, $\gamma_i$, the energy spectrum is found from numerical solution of the dispersion equation~\eqref{disp_eq}.
Figure~\ref{fig1} presents the energy spectrum in 
the SL with the layer widths $l_1=0.3d$, $l_2=0.7d$, and the strain amplitudes $\gamma_1=0.7$, $\gamma_2=-0.3$.
One can see folding of the energy spectrum at the boundaries of the Brillouin zone of the SL ($k=\pm\pi/d$), and the conical dispersion in the direction perpendicular to the strain. 
Figure~\ref{fig1} demonstrates that 
even the large strain with $\gamma \sim 1$ results 
just in small deviations from the pure conical dispersion which occur
in close proximity of the anticrossing points.
Therefore the energy spectrum and the electronic states can 
be described in the nearly-free electron model where the strain is considered as a small perturbation. In this approach, the unperturbed energy spectrum in the $n$th conduction miniband 
is simply given by
\begin{equation}
\label{E_n_0}
	E_n^{(0)}(k,q)= \hbar v_0 \sqrt{q^2+ k_n^2},
\end{equation}
where 
$k_n=k-(2\pi n / d){\rm sign}\{k\}$.
In the following for calculation of absorption in periodically strained graphene we will use 
the perturbation in the form
$$
U(y) =   {\gamma\hbar v_0 \over d} \cos{\left({2\pi y\over d}\right)} \sigma_x 
$$ 
assuming $\gamma \ll 1$. The energies and the wavefunctions in the $n$th miniband are given by
\begin{equation}
\label{E_n}
	E_n = E_n^{(0)} +  \sum_{m\neq n} {|U_{mn}|^2\over E_n^{(0)}-E_m^{(0)}},
\end{equation}
\begin{equation}
\label{psi_perturb}
	\Psi_n = \psi^{(0)}_n + \sum_{m\neq n} {U_{mn}\over E_n^{(0)}-E_m^{(0)}} \psi^{(0)}_m.
\end{equation}
Here $\psi^{(0)}_n$ are unperturbed envelopes
\[	\psi^{(0)}_n = {1\over\sqrt{2}} 
	\left( 
	\begin{array}{c}
	1\\ \pm {\rm e}^{{\rm i}\varphi_n}
	\end{array}
	\right) \, {\rm e}^{{\rm i}q x + {\rm i}k_ny},
\]
where the upper and lower signs are taken for the conduction and valence minibands,  and $\tan{\varphi_n}=k_n/q$.

\section{Dichroic absorption in graphene superlattices}

The Hamiltonian of electron-photon interaction in graphene reads
\begin{equation}
	\label{e-phot}
	H_{e-phot} = - {ev_0\over c} \bm \sigma \cdot \bm A,
\end{equation}
where $\bm A$ is the light-wave vector potential. This yields a general expression for absorbance under direct optical transitions from an initial ($i$) to the final ($f$) miniband:
\begin{equation}
\label{eta}
	\eta_{fi} = 4 {(2\pi e v_0)^2\over c \omega} \sum_{kq} \left|\bm e \cdot \bm \sigma_{fi}\right|^2 \, \delta[E_f(k,q)-E_i(k,q)-\hbar\omega].
\end{equation}
Here the factor 4 accounts for the spin and valley degeneracies, $\omega$ is the light frequency and $\bm e$ is the light polarization vector.

Below in this Section 
we calculate the absorbance for interband transitions in undoped graphene SLs and for transitions between the ground and the first excited conduction minibands for doped SLs. 

\subsection{Interband absorption in undoped SLs}

We consider light absorption 
due to interband transitions between the ground valence and conduction minibands which have opposite dispersions: 
$E_{v0}(k,q)=-E_{c0}(k,q)$. 
In the absence of strain, the interband absorbance has a universal value $\eta = \pi e^2/(\hbar c)$. For normally-incident light, the $D_{2h}$ point group of the symmetrically-strained graphene sheet implies the following polarization-dependence of the absorbance:
\begin{equation}
\label{eta_interband}
	\eta_{c0,v0}(\omega)={\pi e^2 \over \hbar c} \left\{ 1 - \gamma^2 [\Phi_0(\omega) + \Phi_1(\omega) \cos{2\alpha}] \right\}.
\end{equation}
Here $\alpha$ is the angle between the light polarization vector $\bm e$ and $x$ axis.
We take into account in Eq.~\eqref{psi_perturb} only the admixture to the $c0$ and $v0$ states from the closest minibands $c1$ and $v1$, respectively.
\begin{figure}[t]
\includegraphics[width=0.9\linewidth]{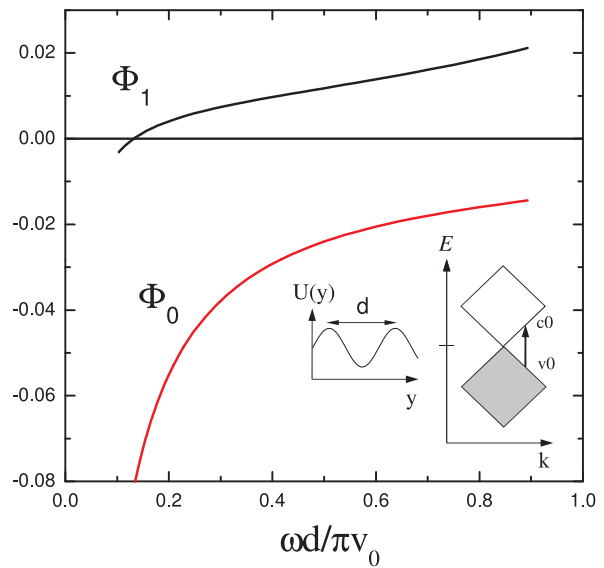}
\caption{Polarization-independent ($\Phi_0$) and polarization-dependent ($\Phi_1$) absorbance spectra under interband transitions in the symmetrical undoped SL. Inset shows the perturbation $U(y)$ caused by periodical strain and the scheme of optical transitions.}
\label{fig2}
\end{figure}
Taking into account the strain 
both in the energies Eq.~\eqref{E_n} and in the optical matrix elements,
we obtain the absorbance in the form of Eq.~\eqref{eta_interband} with
\begin{equation}
	\Phi_0 (\omega)=  {\zeta^2\over 2 \pi^3} \int\limits_0^{\pi/2} {d\varphi} \left[ {4P(4\zeta\sin{\varphi}-1) \over r(r-1)} + \zeta {\partial  P\over \partial \zeta} \right],
\end{equation}
\begin{equation}
	\Phi_1 (\omega) =  {\zeta^2\over 2 \pi^3} \int\limits_0^{\pi/2} {d\varphi} \left( {4P \over r-1} - \zeta {\partial  P\over \partial \zeta} \cos{2\varphi} \right).
\end{equation}
Here $r(\varphi) = \sqrt{1+16\zeta^2-8\zeta\sin{\varphi}}$, 
\begin{equation}
\label{zeta}
P={1 \over r-1}\left(1+ {\cos{2\varphi}+4\zeta\sin{\varphi} \over r} \right),
\quad
	\zeta={\pi v_0\over \omega d} .
\end{equation}

Spectral dependences of polarization-independent and polarization-dependent parts of the absorbance, $\Phi_0(\omega)$ and $\Phi_1(\omega)$, are presented in Fig.~\ref{fig2}. One can see that the values of both parts have an order of $10^{-2}\gamma^2$. This means that, in samples with a large enough strain $\gamma \sim 1$, such modulation of absorbance can be detected experimentally. 

The both absorption parts, $\Phi_{0}(\omega)$ and $\Phi_{1}(\omega)$, should tend to constants at frequences
$\omega =0$ and $\omega  = \pi v_0/d$ due to anticrossing of the energy dispersions. We do not obtain this saturation due to oversimplified approach Eqs.~\eqref{E_n}. However the calculated spectra are correct everywhere beyond the narrow spectral regions $\Delta\omega \sim \gamma \pi v_0/d \ll 1$ near these two points.

%

\begin{figure*}[htb]
\includegraphics[width=0.8\linewidth]{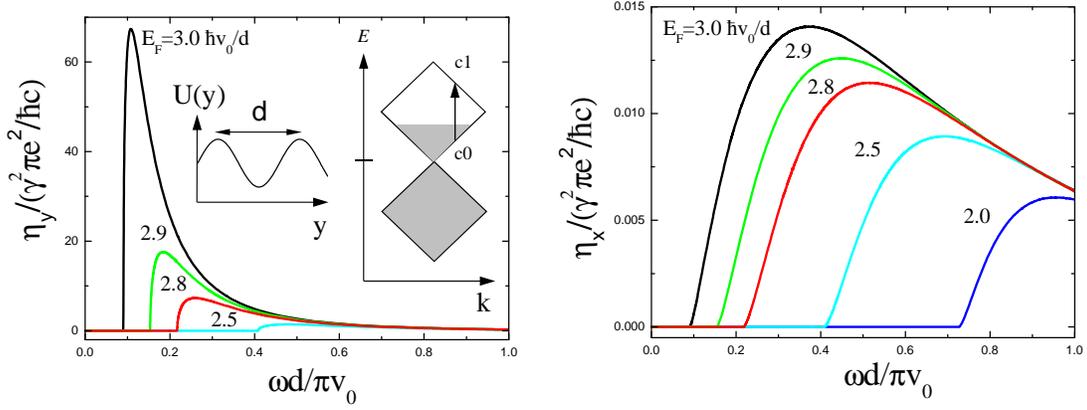}
\caption{Absorbance under inter-miniband transitions in the doped SL. Light polarization is  parallel (left panel) and perpendicular (right panel) to the strain modulation direction. Inset illustrates the scheme of direct optical transitions.}
\label{fig3}
\end{figure*}

\subsection{Inter-miniband absorption in doped SLs}

Let us consider absorption caused by transitions between the ground ($c0$) and the excited ($c1$) conduction minibands. 
The polarization dependence of the absorbance $\eta_{c1,c0}$ can be conveniently written as
\begin{equation}
	\eta_{c1,c0} = \eta_x |e_x|^2 + \eta_y |e_y|^2,
\end{equation}
where $e_x$, $e_y$ are the light polarization vector components perpendicular and along the strain direction, respectively. 

The absorbance is given by
\begin{eqnarray}
	\eta_{x,y} = 4 {(2\pi e v_0)^2\over c \omega} \sum_{kq} \left|\langle c1| \sigma_{x,y}|c0\rangle \right|^2 (f_{c0} - f_{c1}) &&\\
\times	\delta\left[E_1^{(0)}(k,q)-E_0^{(0)}(k,q)-\hbar\omega\right],&&
	\nonumber
\end{eqnarray}
where $f_{c0}$ and $f_{c1}$ are the occupations of the initial and final states. Here we put the unperturbed energies Eq.~\eqref{E_n_0} because the matrix element of inter-miniband transitions already contains $\gamma$.
The calculation yields
\begin{eqnarray}
&& 	\eta_{x,y} = \gamma^2 {4e^2\zeta^2\over \pi^2 \hbar c} \\
	&\times& \int\limits_0^{\pi/2} {d\varphi} (f_{c0} - f_{c1}) 
\sin^2\left( {\varphi-\varphi'\over 2}\right) 
		{s(s+1)\over 2\zeta\sin{\varphi} +1}	
	\, G_{x,y}(\varphi),
	\nonumber
\end{eqnarray}
where $s(\varphi)= (2\zeta^2-1/2)/(1+2\zeta\sin{\varphi})$, $\zeta$ is given by Eq.~\eqref{zeta},
\[	G_x =	\sin^2\left( {\varphi+\varphi'}\right),
	\quad
	G_y = 4\cos^4\left( {\varphi+\varphi'\over 2}\right),
\]
and the angle $\varphi'$ is defined as follows
\[
\cos{\varphi'}= {\cos{\varphi} (s+1)\over s},
\quad
\sin{\varphi'}= {s \sin{\varphi}-2\zeta\over s+1}.
\]

At low temperatures electrons are degenerate, and the light absorption under direct inter-miniband transitions is possible at $E_1>E_{\rm F}>E_0$ only, where $E_{\rm F}$ is the Fermi energy, see inset in Fig.~\ref{fig3}. Therefore the occupation factor is given by $(f_{c0} - f_{c1})  = \theta(E_{\rm F}/\hbar\omega-s)\theta(s+1-E_{\rm F}/\hbar\omega)$, where $\theta(x)$ is the Heaviside function.

The absorbance spectra $\eta_y(\omega)$ and $\eta_x(\omega)$ are presented in Fig.~\ref{fig3}. One can see large dichroism of inter-miniband absorption: $\eta_x \ll \eta_y$. This is explained by different character of electron motion parallel and perpendicular to the strain direction $y$. The light polarized along $x$ can be weakly absorbed because of ballistic electron propagation in this direction. On the other hand, 
the interfaces between strained and unstrained stripes play the role of coherent scatterers, allowing for absorption of the $y$-polarized radiation. 
High absorption dichroism demonstrated here is similar to the result of Ref.~\onlinecite{Dichroism_GNR} where grids of graphene nanoribbons were studied. In contrast, we show that even pure graphene samples can have highly dichroic absorption if they are periodically strained.

The absorbance $\eta_x$ is nonzero in graphene SLs in contrast to conventional semiconductor SLs. The difference is in the electron-photon interaction. According to Eq.~\eqref{e-phot},  the corresponding Hamiltonian in graphene contains the operator $\sigma_x$ while in semiconductor SLs $H_{e-phot}$ is proportional to the unit operator with zero matrix elements between the states in different minibands.

Let us compare  $\eta_{x,y}$ calculated above with the Drude absorbance accompanied by scattering with the characteristic time $\tau$, $\eta_D \sim (E_{\rm F}/\hbar \omega^2\tau) (e^2/\hbar c)$. For realistic parameters $E_{\rm F} \sim \hbar \omega \sim 0.1$~eV, $\tau\sim 1$~ps we have $\eta_D \sim 10^{-2} e^2/(\hbar c)$, which is about an order of magnitude smaller than $\eta_y$ and comparable with $\eta_x$ at $\gamma \sim 1$. This demonstrates that the absorption dichroism caused by the periodic strain is observable in real samples where the Drude absorption is also present.

\begin{figure*}[t]
\includegraphics[width=0.8\linewidth]{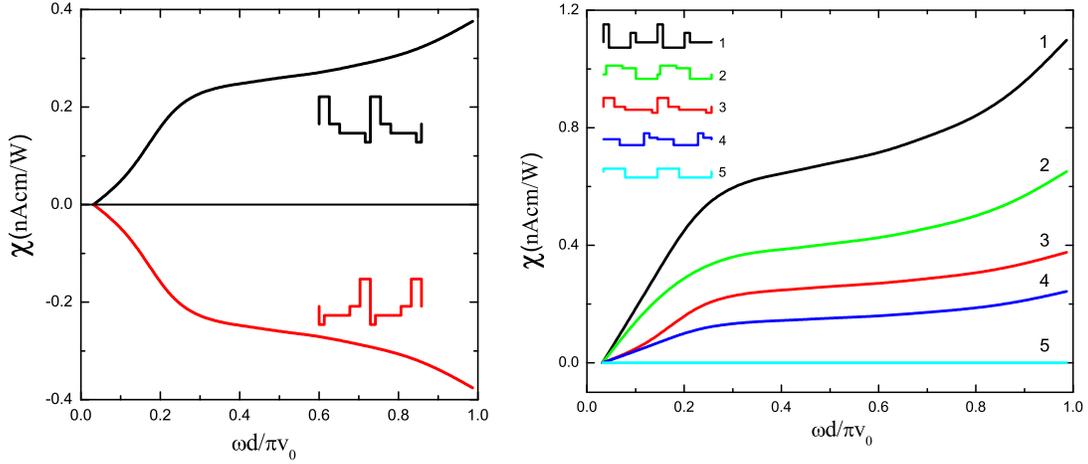}
\caption{Circular-polarization sensitive photocurrent in asymmetrical SLs. Insets show strain profiles for different SLs. The difference between the highest and lowest strain in the layers, $|\gamma_{max}-\gamma_{min}|=0.8, 0.45, 0.5, 0.4$ and 0.3 for the structures 1\ldots 5, respectively. The upper SL in the left panel coincides with the structure \#3 in the right panel. The period $d=10$~nm, the relaxation time in the $c0$ miniband $\tau_0=1$~ps, the Fermi energy $E_{\rm F}=3\hbar v_0/d$.}
\label{fig:CPGE}
\end{figure*}

\section{Circular ratchet photocurrent}

Consider a graphene SL formed by asymmetrical strain lacking the reflection operation $y \to -y$, see insets in Fig.~\ref{fig:CPGE}. 
This system has $C_{2v}$ symmetry with the $C_2$ axis along the $y$ direction. Symmetry considerations imply that absorption of normally-incident circularly polarized light is accompanied by generation of a photocurrent flowing perpendicular to the strain direction. This means that asymmetrical graphene SLs serve as quantum ratchets.

The phenomenological expression for the photocurrent density has the form
\begin{equation}
\label{j_circ}
	j_x = \chi I P_{circ},
\end{equation}
where $P_{circ}={\rm i} (e_x e_y^* - e_x^* e_y)$ is the light circular polarization degree.
Microscopically, the density of the photocurrent generated under inter-miniband optical transitions in doped graphene is given by
\begin{equation}
\label{j}
	j_x = 2 e \sum_{kq,\nu} W_{c1,c0}^{(\nu)} \, (v_x^{(1)}\tau_1-v_x^{(0)}\tau_0).
\end{equation}
Here $\nu$ enumerates the two valleys, $W_{c1,c0}^{(\nu)}$ is the inter-miniband optical transition rate in the $\nu$th valley, $\hbar v_x^{(n)} = \partial E_n/\partial q$, and $\tau_n$ is the momentum relaxation time in the $n$th miniband.
The expression for $\chi$ has  the following form:
\begin{eqnarray}
	\chi = & \dfrac{8\pi^2 e^3 v_0^2}{\omega^2 \hbar c} \sum\limits_{kq,\nu} (v_x^{(1)}\tau_1-v_x^{(0)}\tau_0) \,
	 {\rm Im} \left\{ 
M_x^{(\nu)}{M_y^{(\nu)}}^*
	 \right\} & 
	 \nonumber \\ 
	&\times (f_{c0}-f_{c1}) \, \delta[E_1(k,q)-E_0(k,q)-\hbar\omega],&
	\label{chi}
\end{eqnarray}
where $M_{x,y}^{(\nu)}=\langle \nu c1| \sigma_{x,y} |\nu c0\rangle$.

Numerical calculations performed according to Eq.~\eqref{chi} for various asymmetrical SLs show that 
under the change of the strain in all layers $\gamma_i \to a \gamma_i$
the circular photocurrent scales as
\begin{equation}
\label{chi_gamma}
	\chi \to  a^4 \chi.
\end{equation}
This result can be confirmed by the analysis in the free-electron approximation with a non-centrosymmetric periodic perturbation. For this purpose, let us consider the perturbation of the form
\[
{\cal A}_x = {\hbar\over d} \left[ \gamma_1\cos{2\pi y\over d}+\gamma_2\cos{\left({4\pi y\over d} + \beta\right)} + \gamma_4\cos{8\pi y\over d} \right],
\]
where $\beta \neq 0,\pi$.
It turns out that, for such perturbation, nonvanishing circular photocurrent is obtained only if all the three amplitudes, $\gamma_1, \gamma_2, \gamma_4$, are nonzero.
Indeed, for direct optical transitions between $c0$ and $c1$ minibands, the transferred momentum $2\pi/d$ is taken up by the periodic strain, so the matrix elements
$M_{x,y}^{(\nu)}\propto\gamma_1$. 
In order to take into account the SL asymmetry, one should consider the $\gamma_2$-related contribution. However, the momentum transfer is twice larger in this case, and the correction to the matrix element can be $\delta M_{x,y}^{(\nu)}\propto\gamma_1\gamma_2$ only. The interference of these two corrections yield the $\gamma_1^2\gamma_2$ contribution to ${\rm Im} \left\{M_x^{(\nu)}{M_y^{(\nu)}}^*\right\}$ which results in no electric current after summation over two valleys since they differ by signs of $\gamma$'s. 
Moreover, such $\gamma$-odd photocurrent in each valley would mean the valley current~\cite{VO_PRB} at circularly polarized excitation which is forbidden by symmetry. Therefore such a contribution nullifies already after summation over $q$ and $k$.

The only possibility to get a nonzero photocurrent is the fourth order in the strain. It can be obtained from interference of the first-order term $M_{x,y}^{(\nu)}\propto\gamma_1$ with a  third-order contribution. However it is impossible to construct the correction to the $c0 \to c1$ transition matrix element which $\propto \gamma_1^2 \gamma_2$. The fourth-order in strain probability of the optical transition is obtained as a product of  $M_{x,y}^{(\nu)}$ and $\delta M_{x,y}^{(\nu)} \propto\gamma_1\gamma_2\gamma_4$. This yields the circular ratchet current Eq.~\eqref{j_circ} with  $\chi \sim \gamma_1^2\gamma_2\gamma_4$. This dependence coincides with the numerical result Eq.~\eqref{chi_gamma}.

Figure~\ref{fig:CPGE} presents results of numerical calculation of the circular 
ratchet
photocurrent for various asymmetrical SLs. We assume that the momentum relaxation time in the 1st miniband is much shorter than in the ground miniband, $\tau_1 \ll \tau_0$, and disregard the corresponding contribution to the photocurrent. 
In calculation we take the period $d=10$~nm, the relaxation time $\tau_0=1$~ps, and the Fermi energy $E_{\rm F}=3\hbar v_0/d$.

Left panel of Fig.~\ref{fig:CPGE} represents the circular ratchet current for two strains with opposite asymmetry, i.e. for two SLs transforming one to another by a reflection $y \to - y$. The calculations demonstrate that the circular photocurrents are directed oppositely and have the same magnitude. The right panel of Fig.~\ref{fig:CPGE} shows the circular photocurrents for different strained SLs. One can see a larger photocurrent for SLs with higher degree of asymmetry. For the symmetrical SL (structure \#~5) the circular photocurrent is zero.

\section{Conclusion}

High dichroic absorption in graphene SLs demonstrated above can be used in polarizers operating in infrared and terahertz ranges.
The magnitude of the circular photocurrent is $\chi \approx 1$~nA~cm/W at typical 
photon energy $\hbar\omega \sim \hbar v_0/d \sim 60$~meV for the period $d=10$~nm.
These estimates are close to experimental conditions realized in Refs.~\onlinecite{Karch10_condmat,Jiang11}, which allows one to expect observation of the considered effects in experiments.

To summarize, graphene SLs formed by periodic strain are considered theoretically.
We show that electron energy spectrum consists of minibands obtained by folding of the cone at the boundaries of the SL Brillouin zone with very narrow anticrossing regions.
Light absorption under direct interminiband transitions is shown to be frequency dependent and dichroic. Giant dichroism of absorption is demonstrated for doped  graphene SLs. 
Asymmetrical graphene SLs are shown to act as quantum ratchets allowing for generation of photocurrent at absorption of normally incident light. The helicity-dependent photocurrent spectrum is calculated for doped SLs with various asymmetries.

\acknowledgments
We thank E.L. Ivchenko for stimulating discussions. The work was supported by RFBR, President grant for young scientists, and ``Dynasty'' Foundation -- ICFPM.

\end{document}